\documentclass[aps, prb, superscriptaddress, preprint, onecolumn]{revtex4-1}

\usepackage{graphicx}
\usepackage{xcolor}
\usepackage{bm}
\begin{document}

\title{High-resolution resonant inelastic soft X-ray scattering as a probe of the crystal electric field in lanthanides demonstrated for the case of CeRh\textsubscript{2}Si\textsubscript{2}}

\author{A. Amorese}
\affiliation{European Synchrotron Radiation Facility, 71 Avenue des Martyrs, CS40220, F-38043 Grenoble Cedex 9, France}

\author{N. Caroca-Canales}
\affiliation{Max Planck Institute for Chemical Physics of Solids, N\"othnitzer Strasse 40, D-01187 Dresden, Germany}

\author{S. Seiro}
\affiliation{Max Planck Institute for Chemical Physics of Solids, N\"othnitzer Strasse 40, D-01187 Dresden, Germany}
\affiliation{Leibniz Institute for Solid State and Materials Research, Helmholtzstraße 20, D-01069 Dresden, Germany}

\author{C. Krellner}
\affiliation{Kristall- und Materiallabor, Physikalisches Institut, Goethe-Universit\"at Frankfurt, Max-von-Laue Stasse 1, 60438 Frankfurt am Main, Germany}

\author{G. Ghiringhelli}
\affiliation{CNR-SPIN and Dipartimento di Fisica, Politecnico di Milano, piazza Leonardo da Vinci 32, Milano I-20133, Italy}

%\author{L. Braicovich}
%\affiliation{CNR-SPIN and Dipartimento di Fisica, Politecnico di Milano, piazza Leonardo da Vinci 32, Milano I-20133, Italy}

\author{N. B. Brookes}
\affiliation{European Synchrotron Radiation Facility, 71 Avenue des Martyrs, CS40220, F-38043 Grenoble Cedex 9, France}

\author{D.V. Vyalikh}
\affiliation{Saint Petersburg State University, Saint Petersburg 198504, Russia}
\affiliation{Donostia International Physics Center (DIPC), Departamento de Fisica de Materiales and CFM-MPC UPV/EHU, 20080 San Sebastian, Spain}
\affiliation{IKERBASQUE, Basque Foundation for Science, 48011 Bilbao, Spain}

\author{C. Geibel}
\affiliation{Max Planck Institute for Chemical Physics of Solids, N\"othnitzer Strasse 40, D-01187 Dresden, Germany}

\author{K. Kummer}
\email{kurt.kummer@esrf.fr}
\affiliation{European Synchrotron Radiation Facility, 71 Avenue des Martyrs, CS40220, F-38043 Grenoble Cedex 9, France}

\date{\today}

\begin{abstract}
The magnetic properties of rare earth compounds are usually well captured by assuming a fully localized $f$ shell and only considering the Hund´s rule ground state multiplet split by a crystal electric field (CEF). Currently, the standard technique for probing CEF excitations in lanthanides is inelastic neutron scattering. Here we show that with the recent leap in energy resolution, resonant inelastic soft X-ray scattering has become a serious alternative for looking at CEF excitations with some distinct advantages compared to INS. As an example we study the CEF scheme in CeRh\textsubscript{2}Si\textsubscript{2}, a system that has been intensely studied for more than two decades now but for which no consensus has been reached yet as to its CEF scheme. We used two new features that have only become available very recently in RIXS, high energy resolution of about 30 meV as well as polarization analysis in the scattered beam, to find a unique CEF description for CeRh\textsubscript{2}Si\textsubscript{2}. The result agrees well with previous inelastic neutron scattering and magnetic susceptibility studies. Due to its strong resonant character, RIXS is applicable to very small samples, presents very high cross sections for all lanthanides, and further benefits from the very weak coupling to phonon excitations. The foreseeable further progress in energy resolution will make this technique increasingly attractive for the investigation of the CEF scheme in lanthanides.
\end{abstract}


\maketitle

\section{Introduction}

Resonant inelastic X-ray scattering (RIXS) in the soft X-ray range has seen a rapid development in recent years. As a result, the most modern RIXS spectrometers today can achieve an energy resolution of a few tens of meV at 1 keV incident photon energy, the possibility of truly three-dimensional mapping in \textbf{q} space, and allow for polarization analysis in the scattered beam. This opens new perspectives for studying the low energy excitations in strongly correlated systems at energy scales relevant for the material properties. Recent examples that showcase these new possibilities are the work on orbital, spin and charge-density wave excitations in cuprates and other transition metal oxides.\cite{ghiringhelli2004-prl, moretti2011-njp, schlappa2012-science, chaix2017-nphys, peng2017-nphys, miao2017-pnas, betto2017-prb} This high-resolution RIXS work has focused on strongly correlated, transition metal compounds. By contrast, no work on 4\textit{f} systems has been reported. One of the reasons is that the energy scale of the interesting excitations in the rare-earth intermetallics does usually not exceed 50 meV which has been a major challenge for RIXS. Here we report the first high-resolution RIXS study of low energy magnetic excitations in rare-earth intermetallics and demonstrate the capabilities of this new technique for characterizing the crystal electric field (CEF) in these materials. 

These CEF excitations are a consequence of the broken spherical symmetry when a rare-earth ion is placed in a crystalline environment. In most cases the hybridization of the 4\textit{f} states with the valence states of the surrounding atoms is weak. Then the effect of the surrounding crystal can be treated as an effective electric potential created by the neighboring atoms which is acting on atomic-like 4\textit{f} states. The CEF will split the ground state multiplet of the rare earth ion depending on the point symmetry at the lanthanide sites which results in a huge and important effect on the magnetic properties of lanthanide based compounds. For instance, the large magnetic anisotropy which is frequently observed in lanthanide systems and which is basis for a number of applications is a result of the CEF.

Once the CEF of a system has been well characterized the single ion model usually yields very good agreement with the thermodynamic properties of the system at high and intermediate temperatures. At lower temperatures the effect of intersite interactions, as for example the RKKY exchange interaction, or the coupling of the 4\textit{f} electrons to valence states, e.g. the Kondo interaction, become important and results on the one hand in deviation from the single ion CEF behavior, and on the other hand in a number of phenomena of fundamental relevance, like magnetic ordering, formation of heavy fermions, and quantum critical points. Recently the direct interplay of CEF and 4\textit{f}-conduction hybridization has moved into focus, since it might induce a new kind of transition named “meta-orbital transition”.\cite{hattori2010-jpsj, pourovskii2014-prl, rueff2015-prb} In order to extract the signatures of these exotic physics out of the data, a very good description of the conventional contributions due to CEF effects is needed. For this reason, a wealth of experimental studies dealing with the experimental determination of the CEF scheme can be found in the literature.

To date, inelastic neutron scattering is the standard technique for CEF studies as it can directly probe excitations from the crystal field ground state into excited states and thus obtain information on the  energy splittings between the CEF levels and the symmetry of the levels. We have recently shown that these excitations can in principle also be seen with RIXS at the $M_{4,5}$ edges of the lanthanides.\cite{amorese2016-prb} However, up to now the available energy resolution of 100 meV and more was simply not sufficient for resolving the CEF splittings in rare earth ions (few tens of meV). Here we show that with the energy resolution provided by the most modern spectrometers, RIXS can become a valuable alternative to inelastic neutron scattering with some distinct advantages and disadvantages that make it very complementary to INS as we will discuss below. 

We have performed our exploratory study on CeRh\textsubscript{2}Si\textsubscript{2} as the reported CEF splittings are relatively large (up to 50 meV) and well compatible with the resolution achieved in RIXS to date. CeRh\textsubscript{2}Si\textsubscript{2} shows a very interesting phase diagram with AFM order and unconventional superconductivity in close proximity and therefore attracts a lot of attention.\cite{movshovich1996-prb, araki2002-jpcm, knafo2010-prb, gotze2017-prb, knafo2017-prb} However, the CEF schemes reported in the literature are contradictory both in terms of the observed energy splittings as well as the symmetry and the anisotropy of the CEF levels of the $^2F_{5/2}$ multiplet.\cite{ willers2012-prb, settai1997-jpsj, abe1998-jmmm, patil2016-ncomm} Therefore, a study of the CEF in CeRh\textsubscript{2}Si\textsubscript{2} is also interesting in itself beyond showcasing the capabilities of RIXS.

\section{Experimental details and calculations}

The measurements were performed at the ID32 beamline of the ESRF. This instrument is the first of a new generation of soft X-ray RIXS spectrometers and offers an energy resolution of about 30 meV at the Ce $M_{4,5}$ edges, the possibility of continuously changing the scattering angle allowing 3D \textbf{q} dependent measurements and, in addition, enables polarization analysis in the scattered beam. CeRh\textsubscript{2}Si\textsubscript{2} single crystals have been grown using a standard Czochralski technique in a tri-arc furnace and oriented prior to the experiment using X-ray Laue diffraction.

For the multiplet calculations we have used the Quanty code\cite{haverkort2012-prb, haverkort2016-jpcs}. In this code the CEF is parametrized by weighted coefficients $A_{k,m}$ of the expansion of the crystal field potential onto renormalized spherical harmonics. These coefficients $A_{k,m}$ have to be found experimentally. In $D_{4h}$ symmetry the Hund's rule ground state multiplet $^2F_{5/2}$ of a Ce\textsuperscript{3+} ion will split into three Kramers doublets. For negligible mixing between the ground state multiplet and the higher lying $^2F_{7/2}$ multiplet, i.e. assuming $\Delta_{CEF}\ll \Delta_{SO}\approx 300$\,meV (Stevens' approximation), these three doublets are either linear combinations of $|J_z=\pm5/2\rangle$ and $|J_z=\mp3/2\rangle$ states or pure $|J_z=\pm1/2\rangle$ states
\begin{eqnarray}\nonumber
\Gamma_7^{1} &=& \alpha|\pm5/2\rangle + \sqrt{1-\alpha^2} |\mp3/2\rangle\\ \nonumber
\quad \Gamma_7^{2} &=& \sqrt{1-\alpha^2}|\pm5/2\rangle -\alpha  |\mp3/2\rangle \\\nonumber 
\Gamma_6 &=& |\pm1/2\rangle
\end{eqnarray}
and all CEF parameters $A_{k,m}$ parameters except $A_{20}$, $A_{40}$, and $A_{44}$ will be zero. These three parameters fully define the energy of the three states as well as the mixing $\alpha$ in the $\Gamma_7$ states. The RIXS spectra are then calculated using the usual Kramers-Heisenberg Hamiltonian for RIXS.\cite{ament2011-rmp} The intensity of the (quasi-)elastic line in the experimental and calculated spectra is not well defined and we take it as a parameter that is fitted to match the experiment. In contrast, the energy position and relative intensity of the excitations with non-zero energy are fully defined by the $A_{k,m}$ parameters.

%The energies of these three CEF levels and the mixing factor $\alpha$ of the $\Gamma_7$ states as a function of the $A_{k,m}$ are given by
%\begin{eqnarray}\nonumber
%E_{\Gamma_7^{1,2}} =& - \frac{1}{105} \left(12 A_{20} + 5 A_{40} \pm X\right) \\ \nonumber
%E_{\Gamma_6} =& \frac{8}{35} A_{20} + \frac{2}{21} A_{40} \\ \nonumber
%\alpha =& \sqrt{\frac{175}{X^2-(18A_{20}-10A_{40})X}}\; A_{44} 
%\end{eqnarray}
%with $X = \sqrt{4 (9 A_{20} - 5 A_{40})^2 + 350 A_{44}^2}$.

\begin{table}
\begin{ruledtabular}
\begin{tabular}{l|ccc|ccc}
 & $A_{20}$ & $A_{40}$ & $A_{44}$ & $\Delta_1$ & $\Delta_2$ & $\alpha$ \\\hline
RIXS (this work)   & 95 & 35 & 45 & 30 & 53 & 0.96 \\
INS + XAS  \cite{willers2012-prb} & 64 & 101 & 88 & 30 & 52 & $\pm0.73$ \\
Magn. suscept. \cite{settai1997-jpsj} & 106 & 65 & 33 & 27 & 59 & $\pm0.97$ \\
Magn. suscept. \cite{abe1998-jmmm} & 48 & -174 & 177 & 32 & 80 & $\pm0.90$ \\
ARPES  \cite{patil2016-ncomm} & \textcolor{gray}{85} & \textcolor{gray}{-83} & \textcolor{gray}{120} & 48 & 62 & \textcolor{gray}{$\pm0.93$}\\
\end{tabular}
\end{ruledtabular}
\caption{\label{tab:ceflist} Summary of the crystal field parameters $A_{k,m}$ and splittings $\Delta$\textsubscript{CEF} (both in meV) as well as the mixing $\alpha$ reported for CeRh\textsubscript{2}Si\textsubscript{2}. Note that ARPES gives only information on the splittings, but not on $\alpha$. We have therefore chosen the $A_{k,m}$ (or $\alpha$) such that a good agreement with the magnetic susceptibility reported in Ref. \citenum{settai1997-jpsj} is achieved for the splittings and symmetries reported in Ref. \citenum{patil2016-ncomm} (cf. Fig.~\ref{fig:fig4}).}
\end{table}

\section{Results and discussion}

\subsection{High-resolution RIXS}

\begin{figure}
 \centering
  \includegraphics[width=80mm]{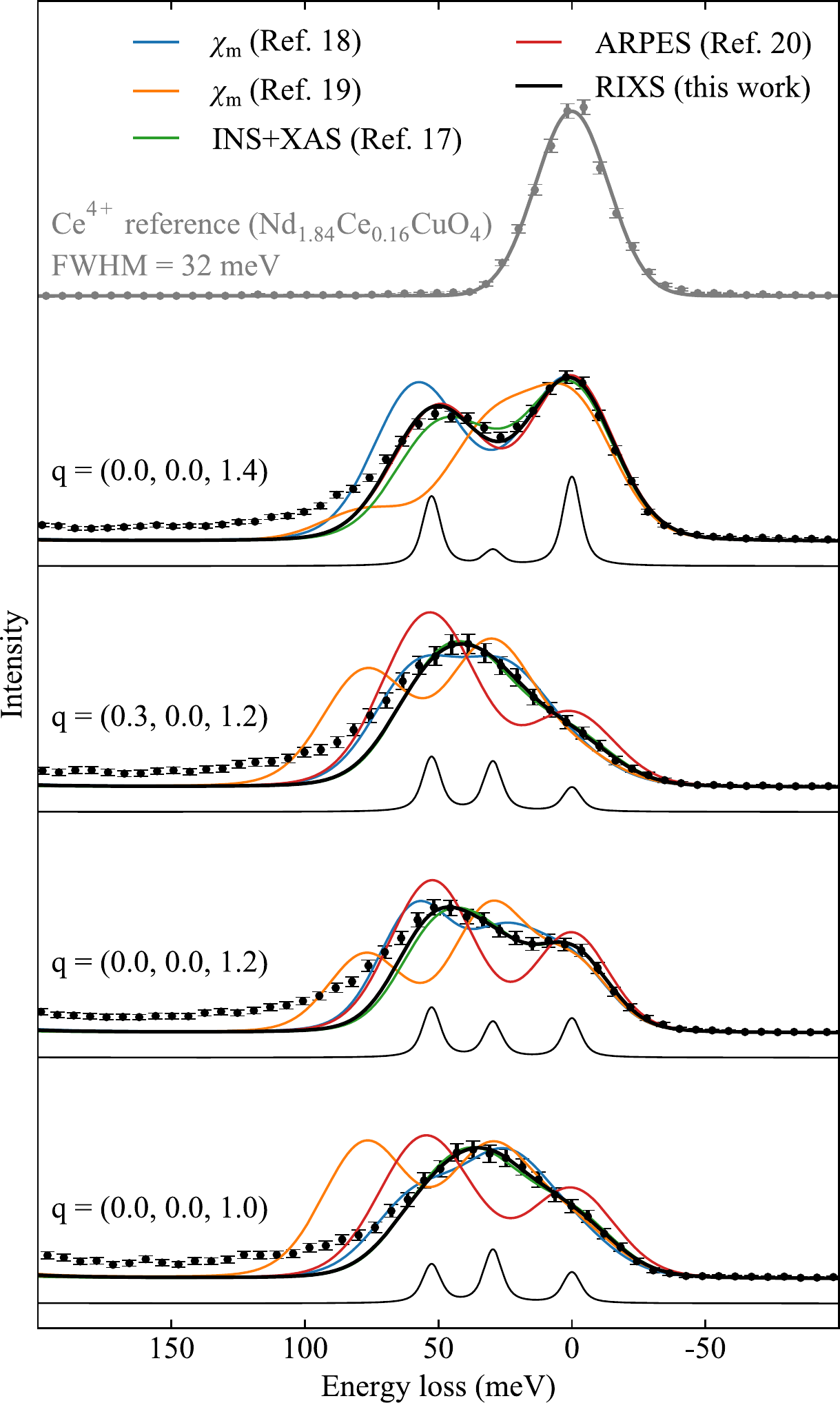}
   \caption{High-resolution RIXS spectra of CeRh\textsubscript{2}Si\textsubscript{2} for several momentum transfers \textbf{q} and incident $\pi$ polarization, compared to calculations broadened with the experimental resolution of 32 meV. In grey the RIXS spectrum of a Ce\textsuperscript{4+} reference sample that is well described by a single, resolution limited Gaussian at zero energy loss.}
\label{fig:fig1}
\end{figure}

In Fig.~\ref{fig:fig1} we show high-resolution RIXS spectra obtained from CeRh\textsubscript{2}Si\textsubscript{2} at various momentum transfers \textbf{q}. The spectra at all \textbf{q} show very good agreement with CEF calculations assuming a $\Gamma_7^1$ ground state with a mixing factor $\alpha=0.96$ and excited levels at 30\,meV and 53\,meV with $\Gamma_7^2$ and $\Gamma_6$ symmetry, respectively (black lines). For comparison we have also included the calculated RIXS spectra for the various CEF schemes proposed in the literature and summarized in Table~\ref{tab:ceflist} (colored lines). Most of the CEF schemes would result in spectra that are in contradiction with our experimental observations at least at some of the measured \textbf{q}'s. In contrast, the scheme previously proposed on the basis of INS+XAS measurements (green lines) is in very good agreement with the RIXS data. This not surprising as the splittings found with RIXS agree very well with those previously obtained with INS (30\,meV and 52\,meV)\cite{willers2012-prb} confirming that RIXS and INS are probing the same magnetic excitations. It should be noted that the collected RIXS spectra are practically free of any phonon contributions. This is because X-rays can only indirectly couple to lattice vibrations via exciton-phonon interaction in the short-lived intermediate state.\cite{devereaux2016-prx} For the strongly screened 3\textit{d}-4\textit{f} exciton this coupling should be very weak and indeed no losses are detected with RIXS when measuring a Ce\textsuperscript{4+} reference where no CEF excitations are present. Hence RIXS provides very clean CEF excitation spectra that can be easily analyzed without the need for additional data from a non-magnetic reference compound.

In order to avoid biasing our analysis we have calculated the RIXS spectra at different \textbf{q} for all possible parameter combinations $(A_{20}, A_{40}, A_{44})$ (-250\,meV $\leq A_{k,m} \leq$ 250\,meV) on a regular, 10\,meV fine mesh and then compared the calculated spectra to the experimental data shown in Fig.~\ref{fig:fig1}. Each parameter set will give a different set of CEF splittings $\Delta_1$, $\Delta_2$, mixing $\alpha$ and symmetry order of the states which fixes the energy position and the relative intensities of the excitations observed in RIXS at a given \textbf{q}. Knowing the energy position and intensity of all three peaks  in the  excitation spectra, the quasi-elastic line at zero energy and the two excitations at $\Delta_1$ and $\Delta_2$, would reduce the possible sets $(A_{20}, A_{40}, A_{44})$ to a few. Unfortunately, the intensity of the (quasi)elastic line at zero energy loss relative to the other, none-elastic features in the RIXS spectra is not well defined. But the lack in information on the intensity of the quasi-elastic line can be compensated by measuring the RIXS spectra at several \textbf{q}, i.e. in different scattering geometries, and using that the intensity ratio between the $\Delta_1$ and $\Delta_2$ excitations varies differently with scattering geometry for different $(A_{20}, A_{40}, A_{44})$.

\begin{figure}
 \centering
  \includegraphics[width=80mm]{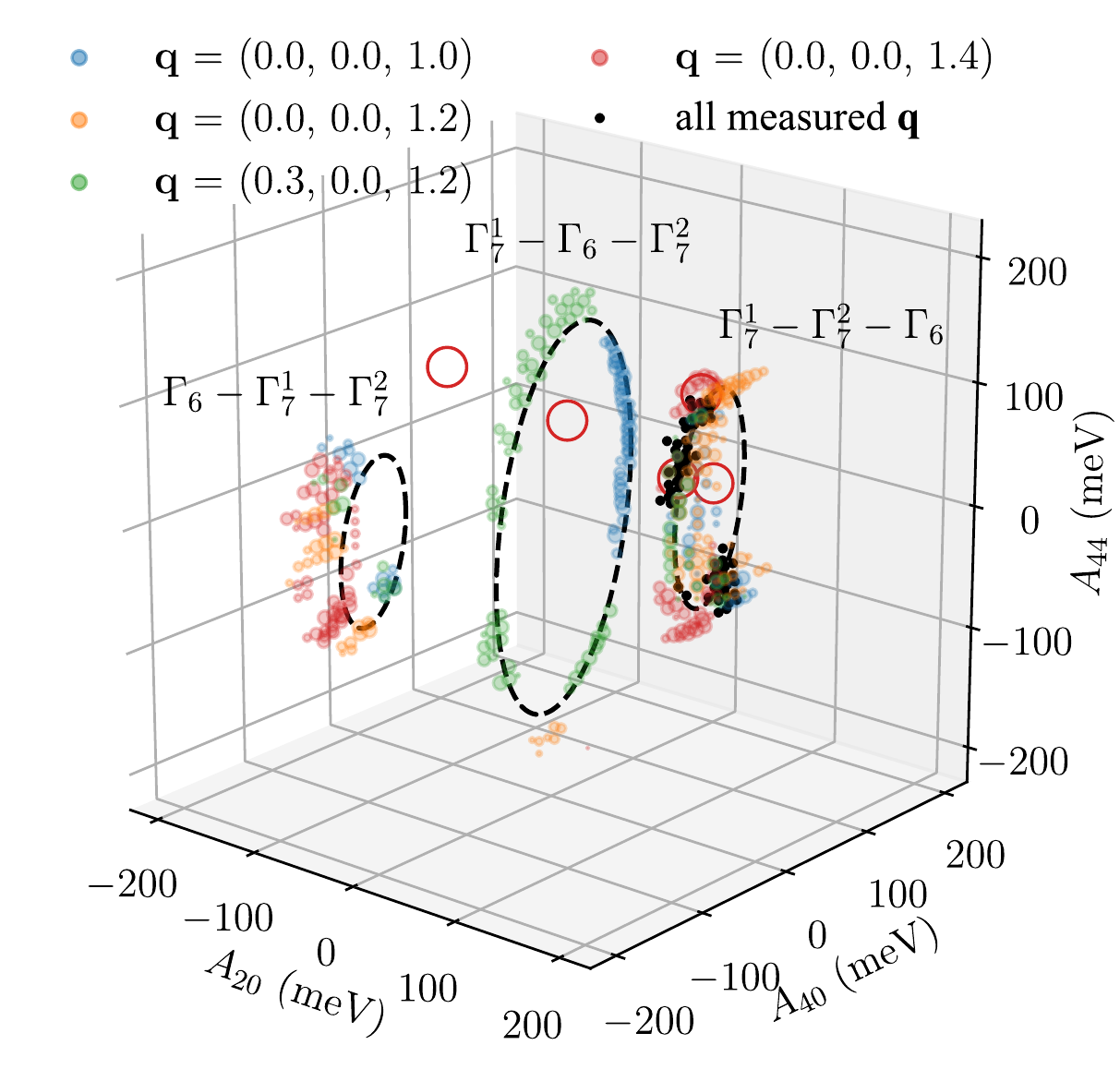}
  \includegraphics[width=80mm]{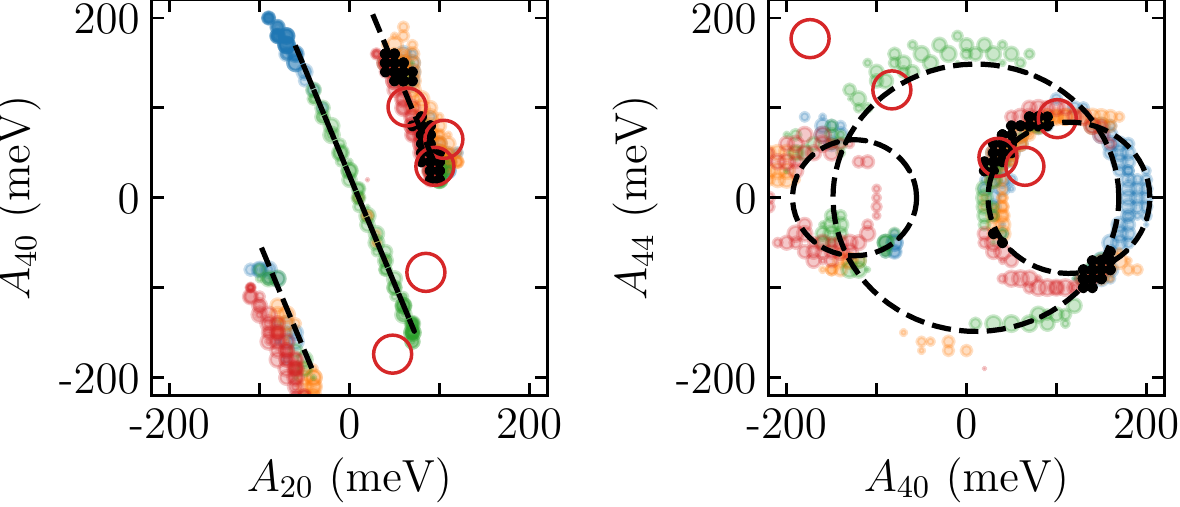}
   \caption{ CEF parameter combinations $\left(A_{20}, A_{40}, A_{44}\right)$ giving a good fit of the RIXS spectra shown in Fig.~\ref{fig:fig1}. Different colors correspond to experimental spectra collected at different \textbf{q}. For each \textbf{q} the 200 points with the lowest $\chi^2$ are shown. The black dots mark those parameter sets that appear for all \textbf{q} and which are therefore  compatible with the RIXS data. The dashed circles show all sets of CEF parameters that yield excitation energies $\Delta_1=30$\,meV and $\Delta_2=53$\,meV. Each circle corresponds to one of the three possible symmetry combinations for the CEF ground and the two excited states. The CEF schemes listed in Table~\ref{tab:ceflist} are marked with red circles.}
\label{fig:fig2}
\end{figure}

The parameter sets giving the best fit for a certain \textbf{q} are shown in Fig.~\ref{fig:fig2} in a different color for each \textbf{q}. The results accumulate around three circles shown with dashed black lines which correspond to those combinations $(A_{20}, A_{40}, A_{44})$ that yield CEF splittings of $\Delta_1=30$\,meV and $\Delta_2=53$\,meV for the three possible orders of the $\Gamma_7^1$, $\Gamma_7^2$, and $\Gamma_6$ levels. Both the radius and the position of these rings will change with $\Delta_1$ and $\Delta_2$ and matching them with the data points is a very robust way of determining the energy splittings. The plot shows that considering the RIXS spectra at only one \textbf{q} can already be enough to extract the energy scale of the CEF excitations. However, in order to get an unambiguous result for the symmetry of the ground and the excited states, i.e. on which of the three circles in Fig.~\ref{fig:fig2} the solution is located, as well as the mixing $\alpha$, i.e. where on that circle the solution is located, one has to combine data at several, different \textbf{q}. Those CEF parameter sets that gave a good fit to the experimental data for all measured \textbf{q} are shown as black dots in Fig.~\ref{fig:fig2}. These points therefore mark the CEF parameter sets that are compatible with the entire data set. We remind again they have been condensed out of all the possible ($A_{20}$, $A_{40}$, $A_{44}$) parameter combinations purely on the basis of the RIXS data shown in Fig.~\ref{fig:fig1} without taking into account any further knowledge on this system other than the point symmetry at the Ce site. Nonetheless we were able to put severe restrictions on the possible CEF scheme in CeRh\textsubscript{2}Si\textsubscript{2}. Clearly the experimental data is only compatible with a $\Gamma_7$ ground state, the first excited state at $(30\pm2)$\,meV, also with $\Gamma_7$ symmetry, and the second excited state at $(53\pm3)$\,meV with $\Gamma_6$ symmetry. This finding is in very good agreement with a previous result obtained with INS\cite{willers2012-prb} showing that both techniques yield compatible information.

\subsection{Temperature dependence and effect of excited CEF levels}

\begin{figure}
 \centering
  \includegraphics[width=80mm]{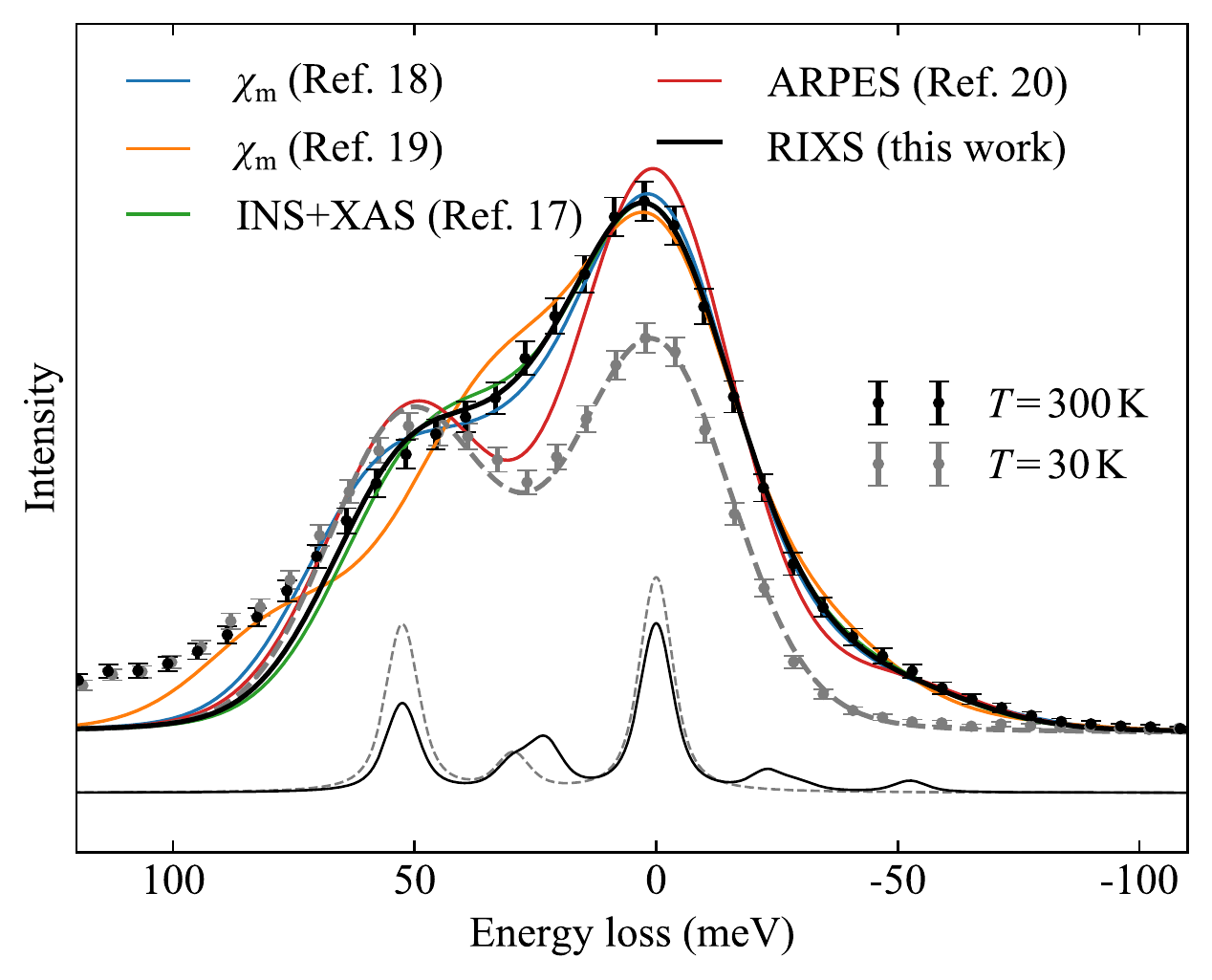}
   \caption{RIXS spectrum of CeRh\textsubscript{2}Si\textsubscript{2} at \textbf{q} = (0.0, 0.0, 1.4) taken at $T=300$\,K compared to calculated spectra for the CEF schemes listed in Table~\ref{tab:ceflist}. Thermal population of excited CEF levels gives rise to additional excitations, including transitions from the excited levels back into the CEF ground state which show up at negative energy loss, i.e. the photons gain energy during the scattering process. For comparison we also show data for $T=30$\,K (gray symbols).}
\label{fig:figHighT}
\end{figure}

The energy scale of excited CEF levels in thermodynamic measurements can be estimated from the observed temperature dependence due to thermal population of excited CEF levels at elevated temperatures. Inelastic scattering experiments, in contrast, are typically performed at temperatures low enough to have only the CEF ground state populated. Then the energy of excited CEF levels is simply given by the position of the peaks in the energy loss spectra. If thermally excited states are involved the expected loss spectra get much more complicated as the number of possible excitations increases and one should, in principle, observe additional losses at energies that correspond to transitions starting from excited CEF levels.

This effect of excited CEF levels on the excitation spectra is shown in Fig.~\ref{fig:figHighT} where we show high-resolution RIXS data collected at \textbf{q}= (0.0, 0.0, 1.4) for $T=30\,$K and $T=300\,$K. The comparison shows that at room temperature the spectral shape of the loss spectrum is significantly different to that at low temperatures. Most notably, excitations at negative energy loss emerge. They correspond to electronic transitions from thermally excited CEF levels back into the CEF ground state, where the excess energy is transferred to the scattered photon. 

We can compare the experimental data at room temperature with multiplet calculations when we start out with an initial state where the $i$th CEF level is populated according to Boltzmann statistics
\begin{eqnarray} \nonumber
P_i = \frac{\mbox{e}^{-\beta\Delta_i}}{\sum_i \mbox{e}^{-\beta\Delta_i}}
\end{eqnarray}
with $\beta=(k_BT)^{-1}$ and $i=0, 1, 2$ running over the three CEF levels of the $^2F_{5/2}$ ground state multiplet.
The resulting spectra are shown as solid lines on top of the data. For the CEF scheme proposed here we find good agreement with the experiment also at room temperature. In contrast, the CEF proposed in Ref.~\citenum{patil2016-ncomm} with excited levels at 48\,meV and 62\,meV (red line) which showed a good agreement with the data at $T=30$\,K (see Fig.~\ref{fig:fig1}) is not compatible with the excitation spectrum at room temperature, both on the energy loss and the energy gain side. The data clearly demonstrates that the expected temperature dependence of RIXS spectra due to thermal population of excited CEF levels (i) can be observed and (ii) contains valuable further information on the energy scale of the CEF excitations.

\subsection{Shape and in-plane orientation of the $\Gamma_7$ wavefunctions: a case for polarization resolved RIXS}

One important aspect of the CEF scheme that has not been addressed yet in our discussion is the mixing $\alpha$ between $|J_z=\pm5/2\rangle$ and $|J_z=\mp3/2\rangle$ in the $\Gamma_7$ states. Besides the energy splitting between the states and their symmetry, this mixing factor is an essential CEF parameter, especially when the CEF ground state is a $\Gamma_7$ states, as in the present case. It then determines the anisotropy at lower temperatures, with $\alpha=0$ or $\alpha=1$ leading to a pure Ising system with large susceptibility along the tetragonal $c$ axis, while $\alpha=\sqrt{3/8}$ results in a pure $XY$ system with a large in-plane susceptibility. Accordingly $\alpha$ dictates the shape of the $\Gamma_7$ wave function. In Fig.~\ref{fig:fig2} the mixing factor $\alpha$ would tell where exactly on the $\Gamma_7^1-\Gamma_7^2-\Gamma_6$ circle the system would be located. The absolute value of $\alpha$ can be determined from magnetic measurements with high accuracy. Therefore, INS results that give the splittings and symmetry of the states are typically combined with measurements of the magnetic susceptibility, or occasionally the linear dichroism in X-ray absorption, in order to fix the mixing and obtain a complete CEF scheme. However, both the anisotropy in magnetic susceptibility and the linear dichroism in XAS depend only on $\alpha^2$ and therefore cannot provide information on the sign of $\alpha$. While the absolute value of $\alpha$ fixes the shape of the $\Gamma_7$ wavefunction, the sign dictates how it is oriented inside the crystal, i.e. whether the lobes are directed towards the (100) or the (110) direction. The sign of $\alpha$ affects for instance the field dependence of the magnetization at high fields. But more importantly, it plays a crucial role in the problem of the meta-orbital transition.\cite{pourovskii2014-prl, rueff2015-prb} It is therefore interesting to ask how much information on the mixing $\alpha$ can be obtained with RIXS only and in combination with other techniques. We will show below that RIXS is very sensitive to the sign of $\alpha$, similar to what has recently been demonstrated for non-resonant inelastic X-ray scattering (NIXS),\cite{willers2012-prl}, as well as the absolute value of $\alpha$.

\begin{figure}
\centering
\includegraphics[width=80mm]{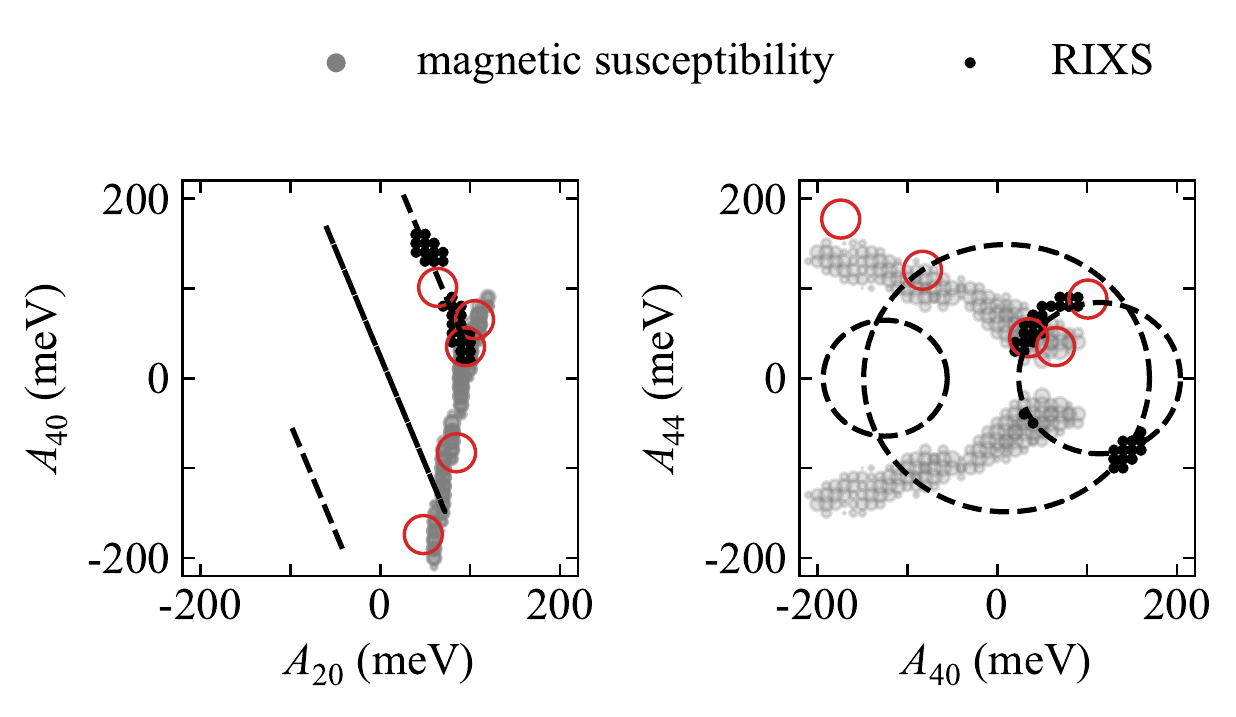}
   \caption{Possible CEF solutions obtained by fitting the RIXS data (black dots in Fig.~\ref{fig:fig2}) and the magnetic susceptibility reported in Ref.~\citenum{settai1997-jpsj}. By combining the information obtained with both techniques and looking where they yield common CEF solutions one can determine both the absolute value and the sign of the mixing factor $\alpha$ with good accuracy. The red circles mark again the CEF schemes listed in Table~\ref{tab:ceflist}.}
  \label{fig:fig_hr_chi}
\end{figure}

It is evident in Fig.~\ref{fig:fig2} that the CEF parameters that give a good fit of the RIXS spectra for all measured \textbf{q} accumulate only in certain sections of the $\Gamma_7^1-\Gamma_7^2-\Gamma_6$ circle. Furthermore the distribution of the data points is not symmetric with respect to the $A_{44}=0$ plane because the RIXS spectra for $+\alpha$ are not the same as those for $-\alpha$. As stated above, each $\alpha$ should, in theory, give a distinct intensity ratio between the three peaks in the $^2F_{5/2}\rightarrow ^2F_{5/2}$ multiplet spectra and one should be able to extract the splittings, the symmetries and $\alpha$ from one single spectrum, but unfortunately the intensities of the experimentally observed as well as the calculated elastic line in RIXS are not well defined. This uncertainty in the intensity of the elastic line and the only slow variation of the intensity ratio of the two loss peaks with $\alpha$ is responsible for the wide distribution of the black dots over almost one eighth of the $\Gamma_7^1-\Gamma_7^2-\Gamma_6$ circle. This can be well seen from Fig.~\ref{fig:fig1} again where the calculated spectra for $\alpha=0.96$ (this work) and $\alpha=\pm0.73$ (Ref. \citenum{willers2012-prb}) both describe the RIXS spectra well while leading to very different magnetic properties of the system. However, because of the asymmetry of the RIXS spectra with respect to the $A_{44}$ plane, the combination of RIXS data with magnetic measurements which are very sensitive to the absolute value of $\alpha$ but are not to the sign of $\alpha$ provides a first, simple way to extract both the absolute value and the sign of $\alpha$. In Fig.~\ref{fig:fig_hr_chi} we plot the limits of the CEF parameters imposed by the RIXS data and those imposed by a fit of susceptibility data. The strong Ising type susceptibility of CeRh\textsubscript{2}Si\textsubscript{2} impose a positive and sizable $A_{20}$, and limits the relative size of $A_{44}$. This combined analysis strongly reduces possible CEF parameters, leaving a very limited range around $A_{20}\approx95$\,meV, $A_{40}\approx40$\,meV, and $A_{44}\approx50$\,meV. Thus combining the information obtained with both techniques one can determine both the absolute value and the sign of all CEF parameters and thus also $\alpha$ with good accuracy.

 \begin{figure}
 \centering
  \includegraphics[width=\textwidth]{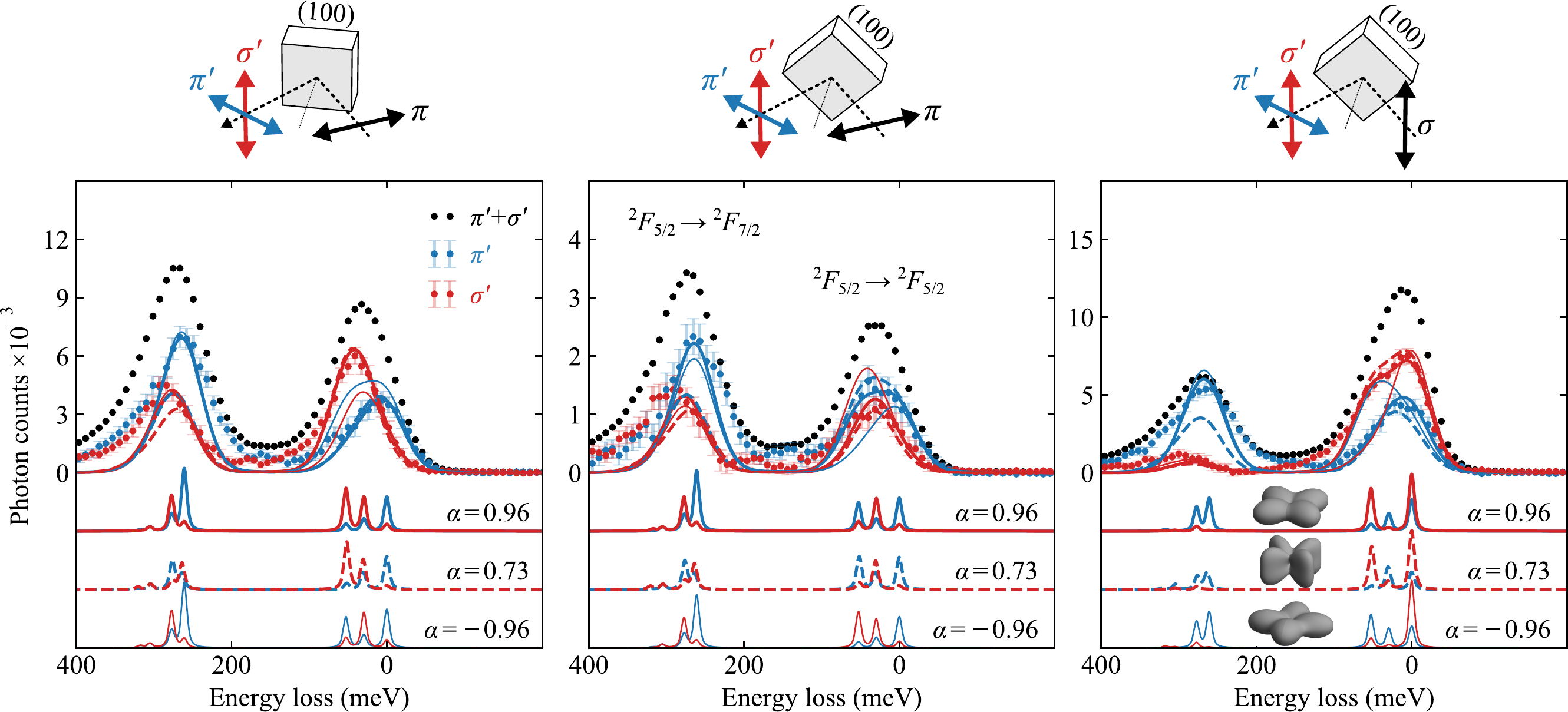}
   \caption{Polarization resolved RIXS spectra taken at \textbf{q} = (0.0, 0.0, 1.2). The incident polarization ($\pi$ or $\sigma$) and the scattering geometry are shown in the sketch above the panels. In black we show spectra obtained without polarisation analysis, in blue and red the decomposition into outgoing $\pi^\prime$ and $\sigma^\prime$ polarization, respectively. The polarization resolved spectra are relatively sensitive to the shape of the wavefunction encoded in the mixing factor $\alpha$. The data are compared to CEF calculations for three different values of $\alpha$ after broadening by the experimental resolution of 55\,meV. We find a good agreement between experimental data and calculations for $\alpha=0.96$ (solid lines) while notable deviations are observed for the lower mixing $\alpha=0.73$ proposed in Ref. \citenum{willers2012-prb} (dashed lines). On the basis of this data one can clearly exclude a negative sign of $\alpha$ which would correspond to a $45^\circ$ rotated ground-state wavefunction around the $C_4$ symmetry axis (thin solid lines). The corresponding ground-state wave functions for each $\alpha$ are shown in the right panel.}
  \label{fig:fig3}
\end{figure}

\begin{figure}
  \centering
   \includegraphics[width=80mm]{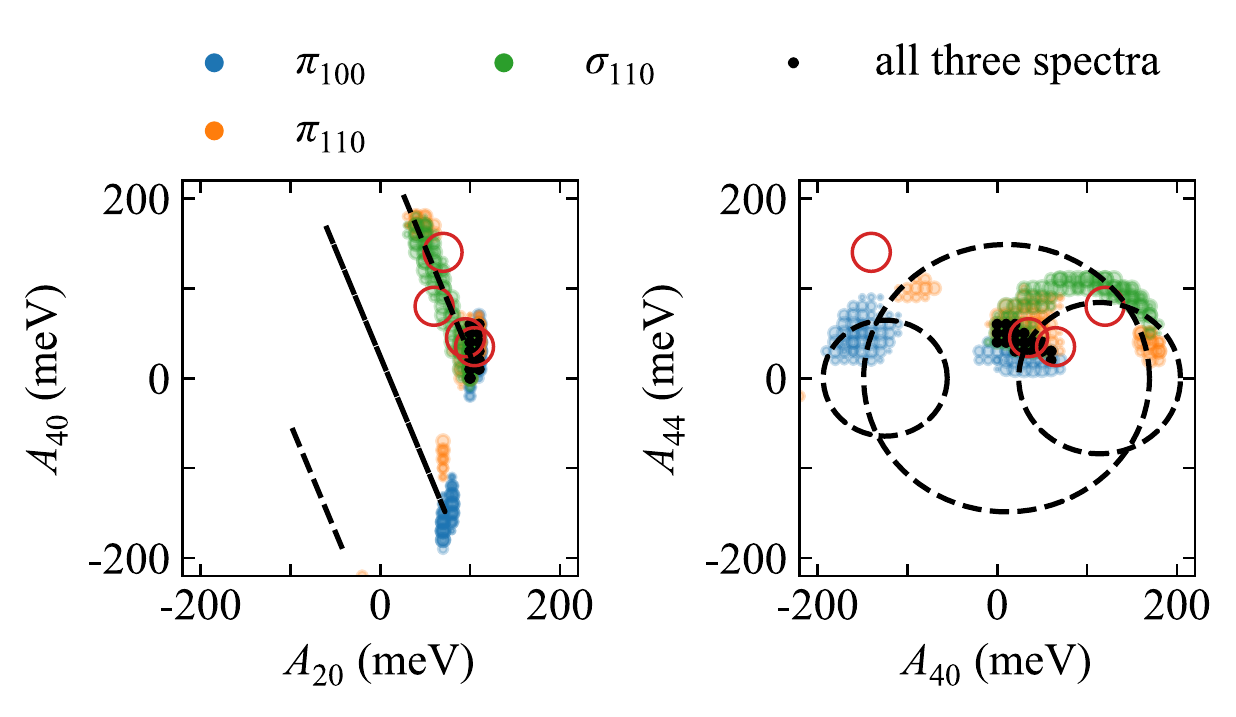}
    \caption{Possible CEF solutions obtained by fitting the three polarization resolved RIXS spectra in Fig.~\ref{fig:fig3}. The points $(A_{20}, A_{40}, A_{44})$ that give the best fit are shown in a different color for each spectrum. The black dots highlight those CEF parameter combinations that emerged for all three spectra. The red circles mark the different CEF schemes listed in Table~\ref{tab:ceflist}.}
\label{fig:figBestAlpha} 
\end{figure}

Polarization resolved RIXS can be a further independent method to get the full information on $\alpha$. While for fixed splittings and symmetry of the CEF levels the RIXS spectral shape without polarization analysis evolves only slowly with $\alpha$, the changes are much more pronounced when separating the crossed and non-crossed polarization components in the scattered beam. This can be used to determine not only the sign but also the absolute value of $\alpha$ with good accuracy directly from RIXS. Soft X-ray RIXS with polarization analysis in the scattered beam has only recently become available \cite{braicovich2014-rsi, minola2015-prl} but is an invaluable tool for the CEF analysis. In Fig.~\ref{fig:fig3} we show polarization resolved RIXS spectra taken at \textbf{q}~=~(0, 0, 1.2) with incident $\pi$ and $\sigma$ polarization and either the (100) direction or the (110) direction of the sample in the scattering plane. Because of the low reflectivity ($\sim$11\%) and efficiency ($\sim$25\%) of the polarizing multilayers used in the soft X-ray range, the data has been collected with the high-throughput configuration of both beamline and spectrometer giving an overall energy resolution of 55 meV.

In order to show the sensitivity of the spectra to the mixing $\alpha$ we compare the experimental data to spectra calculated for $\alpha=0.96$ (this work, solid lines), $\alpha=0.73$ proposed in Ref.~\citenum{willers2012-prb} on the basis of XAS measurements (dashed lines), and $\alpha=-0.96$ (thin solid, lines). All three values of $\alpha$ give a very different response in the two polarization channels, also for the $^2F_{5/2}\rightarrow^2F_{7/2}$ excitations at higher energies, with $\alpha=0.96$ being by far the most compatible with the experimental data for all three combinations of polarization and sample orientation. On the basis of the polarization resolved RIXS data we can therefore determine both the sign and the absolute value of $\alpha$ with much higher precision than from high resolution data shown in Fig.~\ref{fig:fig1} only. In Fig.~\ref{fig:figBestAlpha} we show those parameter sets $(A_{20}, A_{40}, A_{44})$ which best fit the polarization resolved RIXS spectra in Fig.~\ref{fig:fig3}, in a different color for each spectrum. The black dots mark those parameter sets that fit all three spectra and set narrow limits on $\alpha$. The found mixing agrees very well with that obtained from combining the high-resolution data with information from magnetic susceptibility measurements (see Fig.~\ref{fig:fig_hr_chi}). The data clearly demonstrates that RIXS can be used in a similar fashion as NIXS for the determination of the in-plane orientation of orbitals that is given by the sign of $\alpha$, while also giving direct information on the excited CEF levels.

\subsection{Comparison with other techniques}

 \begin{figure}
  \centering
   \includegraphics[width=80mmh]{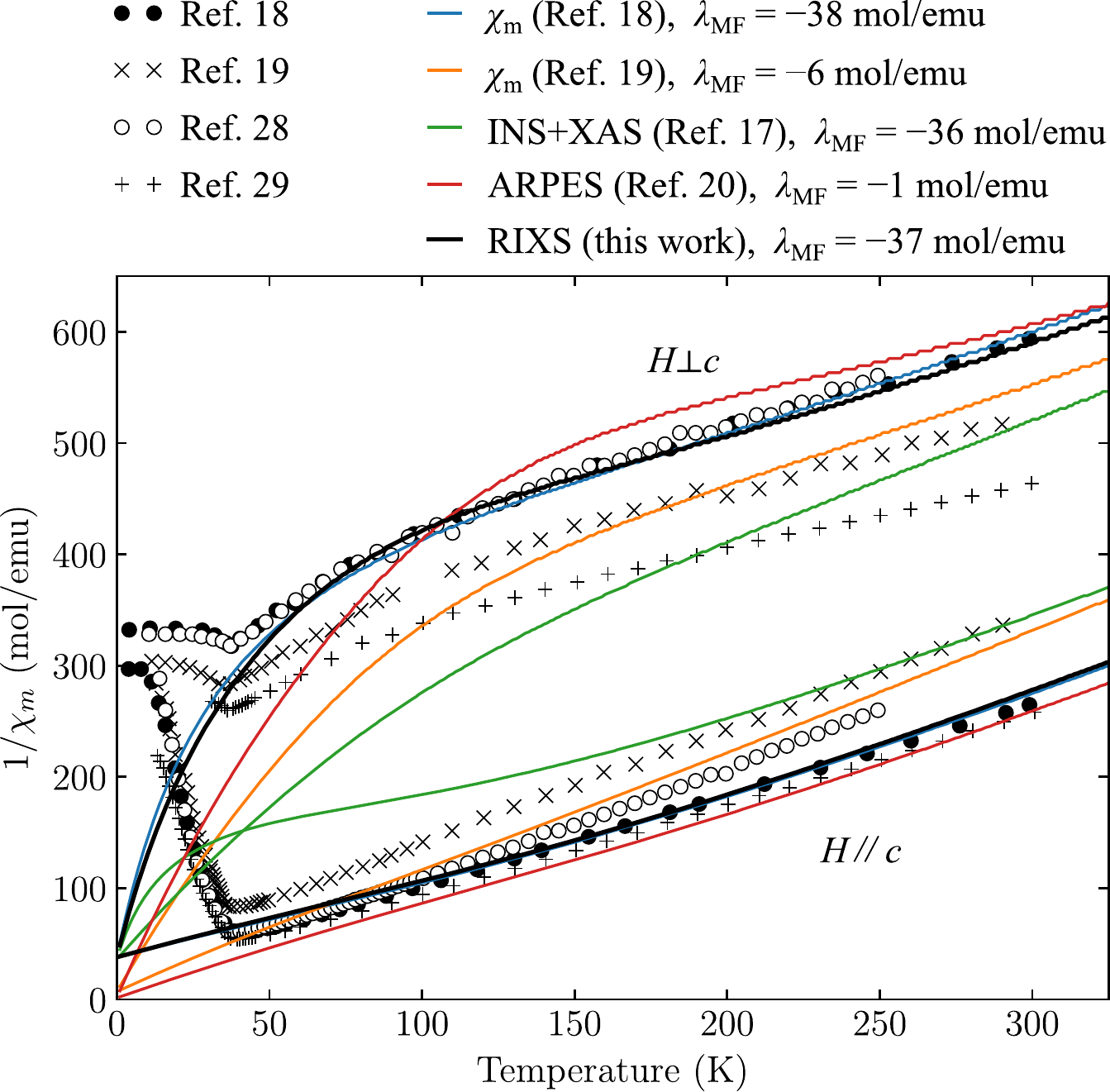}
    \caption{Magnetic susceptibility data reported in the literature compared to the calculated $\chi^{-1} = \chi_{CEF}^{-1}-\lambda_{MF}$ for the CEF parameters listed in Table~\ref{tab:ceflist}. The curves with large absolute values of $\chi^{-1}$ (i.e. small susceptibility) correspond to the in-plane susceptibility, while the curves with the small values of $\chi^{-1}$ (large susceptibilities) corresponds to the $c$-axis susceptibility}
\label{fig:fig4} 
\end{figure}

A good test of the CEF scheme obtained with RIXS is in the comparison to thermodynamic measurements, like magnetic susceptibility or specific heat. The magnetic susceptibility for CeRh\textsubscript{2}Si\textsubscript{2} that is reported in the literature\cite{settai1997-jpsj, abe1998-jmmm, sakai2013-jkps, mori1999-physica} is shown Fig.~\ref{fig:fig4}. While the absolute values for each directions show some scatter between different measurements, all results agree on a huge Ising type anisotropy, i.e. the susceptibility along $c$ is much larger than in the basal plane. For comparison we show the calculated $\chi_m$ as a function of temperature for the different CEF schemes in Table~\ref{tab:ceflist}. The magnetic properties predicted on the basis of the CEF scheme established purely from the RIXS data shown here (black lines) shows a very good agreement with the experimental data reported in Ref.~\citenum{settai1997-jpsj}, even comparable to the agreement of the CEF scheme established by the authors themselves purely on the basis of their magnetic measurements.

In contrast, the mixing factor $\alpha$, proposed previously based on a combined INS and XAS analysis\cite{willers2012-prb} is not only in disagreement with the scheme we deduced from our RIXS data, but also incompatible with the susceptibility data. The reduced $\alpha=0.73$ would result in a much weaker anisotropy, which even changes sign ($\chi_{ab}>\chi_{c}$) below 50 K (green lines in Fig.~\ref{fig:fig4}), in clear contradiction with experiment. Ref.~\citenum{willers2012-prb} argued with a strong exchange anisotropy of opposite sign, which is however in contradiction with the very weak anisotropy observed in the homologue GdRh\textsubscript{2}Si\textsubscript{2} where CEF effects are absent because the 4\textit{f} moment is a pure spin state $J = S = 7/2$.\cite{kliemt2017-prb} Furthermore the proposed CEF ground state with $\alpha=0.73$ bears a sizable saturation moment of $0.93\,\mu_B$ in the basal plane, but a much smaller saturation moment of $0.54\,\mu_B$ along the $c$ axis. This is in complete contradiction with the magnetic structure determined by neutron scattering studies, which find an ordered moment of $1.38\,\mu_B$ pointing along the $c$ direction.\cite{kawarazaki2000-prb} A much smaller ordered moment deduced from NMR results had been reported, but it was demonstrated that this small moment is an artifact resulting from neglecting the long range character of the RKKY interaction.\cite{sakai2013-jkps} It was also argued with the presence of a large exchange field in the AFM state resulting in a larger $\alpha$ and thus in a larger saturation moment along $c$. However, the INS and XAS data based on which the CEF with reduced $\alpha$ was proposed had been taken at 5K, i.e. far in the AFM regime and therefore in the presence of the exchange field. Furthermore, a visible change in the INS and XAS spectra was not observed between the AFM state and the paramagnetic state, which excludes that the exchange field has a strong effect on the CEF scheme. Therefore, the strong deviations between the mixing $\alpha$ deduced from the linear dichroism (LD) in XAS and the observed magnetic susceptibility remain unexplained.

\begin{figure}
  \centering
   \includegraphics[width=80mm]{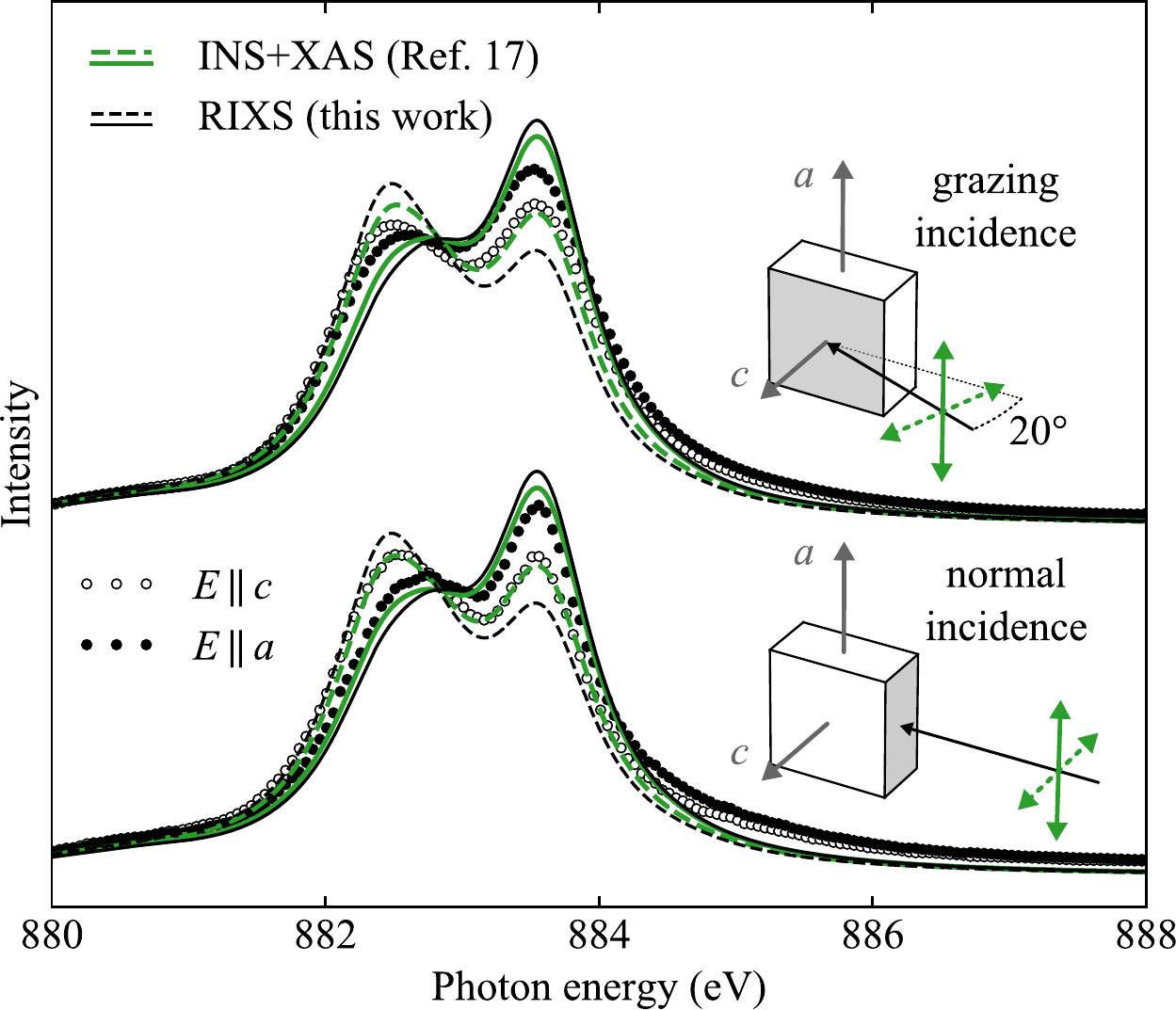}
    \caption{Linear dichroism in the Ce $M_5$ X-ray absorption spectra of CeRh\textsubscript{2}Si\textsubscript{2} observed at $T=300\,$K for grazing and normal incidence. The normal incidence has been taken from Ref.~\citenum{willers2012-prb}, the grazing incidence data has been obtained from a freshly cleaved $ab$ surface. The reduction of the linear dichroism in the grazing incidence spectrum for $E\parallel c$ which is due to the $20^\circ$ angle between the $E$ vector of the light and the $c$ axis of the sample has been corrected. Therefore the amplitudes of the dichroism in the spectra can be directly compared. The lines show the calculated spectra for the CEF parameters reported in Ref.~\citenum{willers2012-prb} (green) and obtained with RIXS (black).}
\label{fig:figXAS} 
\end{figure}

Our study of the polarization dependence of the RIXS signal confirms the CEF ground state in CeRh\textsubscript{2}Si\textsubscript{2} to be an almost pure $|J_z=5/2\rangle$ state, as initially deduced from susceptibility data\cite{settai1997-jpsj, abe1998-jmmm} and compatible with the magnetic moment found in neutron diffraction. 
%But it contradicts the larger mixing of $|J_z=5/2\rangle$ and $|J_z=5/2\rangle$ found linear dichroism (LD) in XAS studies.  
It therefore seems that, at least in the case of CeRh\textsubscript{2}Si\textsubscript{2}, the LD in XAS analysis severely overestimates the mixing of different $|J, J_z\rangle$ states. A possible explanation for this might be in the surface sensitivity of the technique. Soft X-ray XAS data at the $M_{4,5}$ edges of the rare-earth are collected in total electron yield mode for which the probing depth is very short compared to, for instance, the penetration depth of the incoming X-rays (few nm vs ~100\,nm and more). However, close to the surface the CEF acting on the 4\textit{f} site is often modified due to the broken symmetry and the surface relaxation of the crystal lattice. A strong indication for this are the very different CEF splittings $\Delta_1$, $\Delta_2$ observed with surface sensitive photoemission\cite{patil2016-ncomm} compared to what is found with bulk sensitive INS\cite{willers2012-prb} or RIXS (see Table~\ref{tab:ceflist}).

The LD in XAS will be sensitive to these CEF modifications at the surface, too, as a large portion of the signal is coming from the first few unit cells. A strong indication for this sensitivity is given by the data shown in Fig.~\ref{fig:figXAS}. There we compare the LD dichroism of CeRh\textsubscript{2}Si\textsubscript{2} measured once in grazing incidence geometry, where the surface contributions are very high, and once in normal incidence with reduced surface contributions. The small geometric reduction of the LD for grazing incidence by a factor $\sin^2 \theta$ with $\theta=20^\circ$ the incidence angle with respect to the surface plane has been corrected. Therefore, the normal and grazing incidence spectrum should display a comparable LD while experimentally a notably reduced LD and different overall spectral shape is observed for the grazing incidence spectrum which has large surface contributions. But even at normal incidence a significant portion of the XAS signal is still coming from the first few unit cells. Therefore, one can expect that a purely bulk-derived spectrum would shown an even larger LD, in line with what has been calculated for the CEF deduced with RIXS. As the LD is used to determine the mixing $\alpha$, the too large mixing obtained with XAS compared to bulk-sensitive RIXS or susceptibility measurements could be explained by modifications of the CEF close to the surface and the sensitivity of XAS to that. Unfortunately, using bulk-sensitive fluorescence yield detection for the XAS measurements is not an option in the soft X-ray range because strong self-absorption effects result in severely distorted spectral shapes. Therefore, the surface sensitivity of TEY-XAS is an intrinsic problem for CEF work that cannot be overcome easily and that could impair the reliability of LD XAS for CEF studies.

\section{Conclusions}

In summary, using the example of CeRh\textsubscript{2}Si\textsubscript{2} we have shown that RIXS is a powerful tool to study CEF excitations in rare-earth intermetallics. High-resolution RIXS provides information on the energy splittings and the symmetry of the CEF levels. In that regard it is very comparable to INS. But unlike INS, RIXS does not suffer from a phonon background that needs to be characterized. Furthermore, it can be applied to very small, sub-millimeter sized samples and even thin films. RIXS is also sensitive to the shape and orientation of the wave-function of the CEF ground and excited states. It therefore allows to determine both the absolute value and the sign of the mixing angle $\alpha$ with high accuracy, either in combination with magnetic measurements or on its own by using polarization analysis in the scattered beam. The technique is therefore capable of providing complete and unambiguous information on the CEF which usually can only be obtained by combining several different techniques.

As the $M_{4,5}$ resonances are strong across the entire rare earth series, RIXS will be equally suitable to study heavier rare-earth elements including strong neutron absorbers. The cross section and energy resolution in RIXS is preserved over many eV energy loss which allows to measure not only excitations within the ground state multiplet but complete multiplet spectra in the heavier 4\textit{f} elements which show losses up to 10-20 eV. The resonant character of the technique provides with chemical selectivity by tuning the energy to the absorption edge of a particular element. As a photon-in photon-out technique RIXS is truly bulk-sensitive and compatible with the application of magnetic fields which is an important tuning parameter for the study of complex phenomena in rare earth intermetallics. 

On the downside, the technique is not compatible with the application of, for instance, hydrostatic pressure as the employed diamond anvil cells are not transparent to soft X-rays. Most importantly, the energy resolution even of the best RIXS spectrometers today is far from being competitive with what is routinely achieved in INS. Unfortunately, in the employed grating spectrometers it further degrades when going to higher incident photon energies. For the ID32 spectrometer at the ESRF, for instance, the 30\,meV achievable at the Ce $M_5$ edge ($\hbar\omega_{in}\approx880\,$eV) will already be reduced to 50\,meV at the Gd $M_5$ edge (1180\,eV).

Further improvements in the resolution towards 20~meV at 1~keV incident photon energy should be achieved in the near future and make RIXS more suitable also for studies of heavier rare-earth elements. Often hybridization effects broaden the CEF excitations and affect their lineshapes. Therefore, in strongly hybridized cases resolution is often not the limiting factor anymore. The moderately hybridized CeRh\textsubscript{2}Si\textsubscript{2} already shows natural linewidths of about 15-20 meV in INS\cite{willers2012-prb}. For more strongly hybridized cases with even broader excitation, INS can struggle to separate the magnetic excitations from the phonon background. In particular in these cases, the absence of background in RIXS can allow to obtain clean spectra of the CEF excitations. It should be noted, however, that in the presence of strong hybridization effects the simple CEF model employed here usually does not give an appropriate description of the observed excitation spectra as it does not account for the band character acquired by the 4\textit{f} states. More advanced models beyond a single-ion description then become more appropriate.\cite{goremychkin2018-science}

%\section{Conclusion}
%\begin{enumerate} 
%\item high resolution RIXS sees CEF splitting and it is sensitive to symmetry. 
%\item Thanks to the comparison with simulations, we can completely determine the crystal field parameters with no ambiguity.
%\item the use of polarimeter make peaks more visible and gives info about orientation.
%
%\item cf model is limited, in particular for strongly hybridized (mixed-valent) systems
%\item chemical selectivity, small samples ...
%\item (almost) no phonons
%\item high field compatible, but not conventional high pressure
%\end{enumerate} 
%

%\begin{itemize}
	%\item Details of the calculations. Intensities and positions of peaks are fixed except elastic. Assuming sharp levels which is an approximation. Broadening done with experimental resolution.
  %\item How did we get the splitting? Fig.2 alpha is fixed within a certain range but not well defined
  %\item polaristion analysis can do that. very sensitive to alpha and its sign
  %\item comparison with susz. data
%\end{itemize}

\begin{acknowledgments}
We would like thank O. Isnard, G. van der Laan, L.H. Tjeng and A. Severing for valuable discussions, I.P. Makarova for providing us with the Ce\textsuperscript{4+} reference sample, and L. Braicovich for his work on the polarimeter and help with using it. This work was supported by the German Research Foundation (DFG) through grants grant GE602/4-1, KR3831/5-1 and Fermi-NEst. D.V.V. acknowledges support by Saint-Petersburg State University under Research Grant No. 15.61.202.2015. 
\end{acknowledgments}

%\bibliography{biblio}


\end{document}